\documentclass[12pt,a4paper]{iopart}

\usepackage{iopams}
\usepackage{amsthm}
\usepackage{color}
\usepackage[utf8]{inputenc}
\usepackage{graphicx}
\usepackage{graphpap}
\usepackage{cite}
\usepackage{multirow}

\def\eqref#1{\textup{(\ref{#1})}}

\def\defn#1{{\bf #1}}

\def\Cmpx{{\mathbb C}}
\def\Intg{{\mathbb Z}}

\def\innerprod(#1,#2){{\left<#1\,,\,#2\right>}}
\def\Set#1{{\left\{#1\right\}}}
\def\qquadtext#1{\qquad\rm{#1}\qquad}
\def\qquadand{\qquadtext{and}}

\def\ppfrac#1#2{\frac{\partial^2 #1}{\partial {#2}^2}}

\def\F{\mathcal{F}}

\def\Re{\rm{Re}}
\def\Im{\rm{Im}}

\def\Ve{{\bi e}}
\def\Vk{{\bi k}}

\def\VE{{\bi E}}
\def\VB{{\bi B}}
\def\VD{{\bi D}}
\def\VH{{\bi H}}

\def\Vx{{\bi x}}

\def\VP{{\bi P}}

\def\FT#1{\widetilde #1}
\def\FTt#1{\hat{#1}}
\def\FtE{{\hat{E}}}
\def\FtP{{\hat{P}}}

\def\tE{{\widetilde{E}}}
\def\tP{{\widetilde{P}}}
\def\tL{{\widetilde{L}}}

\def\twopii{2\pi i}

\def\Odd{{\textbf{o}}}
\def\Evn{{\textbf{e}}}

\def\omo{{\Omega}}
\def\omodd{\omega^\Odd}
\def\omeven{\omega^\Evn}

\def\oddpn_#1{P^\Odd_{n,#1}}
\def\evenpn_#1{P^\Evn_{n,#1}}

\def\voddpnn#1{\underline{P}^\Odd_{#1}}
\def\vevenpnn#1{\underline{P}^\Evn_{#1}}

\def\pn{P^{n}}

\def\FOmega{{\cal F}}

\begin{document}
\title{Spatially Dispersive Inhomogeneous Electromagnetic Media with Periodic Structure}
\author{Jonathan Gratus\footnote{Corresponding author} and Matthew McCormack}
\address{Physics Department, Lancaster University, Lancaster LA1 4YB and the Cockcroft Institute}
\ead{\mailto{j.gratus@lancaster.ac.uk}, \mailto{m.mccormack@lancaster.ac.uk}}

\begin{abstract}
  Spatially dispersive (also known as non-local) electromagnetic media
  are considered where the parameters defining the permittivity
  relation vary periodically. Maxwell's equations give rise to a
  difference equation corresponding to the Floquet modes.  A complete
  set of approximate solutions is calculated which are valid when the
  inhomogeneity is small.  This is applied to inhomogeneous wire
  media.  A new feature arises when considering spatially dispersive
  media, that is the existence of coupled modes.
\end{abstract}


\submitto{\JOA}

\maketitle

\section{Introduction}

The standard method of calculating the Bloch modes for a given lattice
is to prescribe the constitutive relations for each point within the
unit cell. The corresponding dispersion relations for the bulk material
are then calculated either theoretically or numerically. One may
interpret this dispersion relation in terms of an anisotropic
permittivity and permeability $\FT{\epsilon}_{\rm{Bulk}}$ and
$\FT{\mu}_{\rm{Bulk}}$ for a homogeneous medium. These, in general, are
temporally and spatially dispersive.
However, usually the permittivity and permeability $\FT{\epsilon}$ and
$\FT{\mu}$ of the materials used to construct the unit cell are
considered to be only temporally dispersive, that is, they depend on
the frequency $\omega$ \footnote{Note that in this paper we use $\omega$ to denote the temporal frequency rather than the angular frequency for which it is commonly used.}. Up to now very little theoretical research
has considered periodic media where the materials inside the unit
cell itself are spatially dispersive, i.e. $\FT{\epsilon}(\omega,\Vk)$
and $\FT{\mu}(\omega,\Vk)$, where $\Vk$ is the wave number. A
spatially dispersive medium responds not only to a signal at a
particular point but to signals in the neighbourhood of that point.
For this reason it is also known as a non-local medium. 

The use of a spatially dispersive inhomogeneous medium enables one to
shape the propagation modes, extending the possibilities from the usual
sinusoidal shape.  This may have a number of advantages.  For example
in an accelerating drift tube, one could flatten the mode shape to
enable higher acceleration for a given peak field.  By contrast in
signal transmission one may desire a higher peak for a given total
energy.

In order to
construct an inhomogeneous unit cell it is necessary to consider a
medium where the permittivity and permeability depend on both the wave
vector $\Vk$ and the position $\Vx$ within the unit cell,
i.e. $\FT{\epsilon}(\omega,\Vk,\Vx)$ and
$\FT{\mu}(\omega,\Vk,\Vx)$. In order to make sense of the arguments of
$\FT{\epsilon}(\omega,\Vk,\Vx)$ we no longer interpret
$\FT{\epsilon}(\omega,\Vk,\Vx)$ as the Fourier transform of the
permittivity response function $\epsilon(t,\Vx)$ but in terms of a
differential equation \cite{costa2012macroscopic}. This is given in
equation (\ref{Intr_P_PDE}) below. An alternative interpretation of
$\FT{\epsilon}(\omega,\Vk,\Vx)$ is in terms of a susceptibility kernel
\cite{gratus2010covariant,gratus2011covariant}.

This approach is in contrast with most theoretical articles which
consider inhomogeneous spatially dispersive media. In \cite{
  pekar1958theory, henneberger1998additional,
  maslovski2010generalized, silveirinha2006additional,
  silveirinha2008additional, silveirinha2009additional,
  agranovich1967spatial,agarwal1974electromagnetic,%
  zeyher1972spatial} the inhomogeneity
is restricted to looking at two homogeneous media which are connected
and investigating the corresponding additional boundary conditions.

For this article we will assume that there is a linear constitutive
relationship between the polarization field $\VP(t,\Vx)=\VD(t,\Vx)-
\epsilon_0 \VE(t,\Vx)$ and electric field $\VE(t,\Vx)$. To simplify
the analysis we assume: there is no magnetization so that $\VH=
\mu_{0}^{-1}\VB$ and that all fields are functions of time $t$ and one
spatial coordinate $x=x_1$, thus independent of $x_2,x_3$.  In
frequency domain, $k_2=k_3=0$ and we set $k=k_1$.  The electric and
polarization fields are either linearly polarized transverse modes or
longitudinal modes. For transverse modes $\VE(t,x)=E(t,x)\Ve_2$,
$\VP(t,x)=P(t,x)\Ve_2$ and $\VB(t,x)=B(t,x)\Ve_3$ in the
$(\Ve_1,\Ve_2,\Ve_3)$ frame.  These assumptions automatically
satisfies the two non-dynamic source free Maxwell's equations
$\nabla\cdot \VB=0$ and $\nabla\cdot\VD =0$ and the remaining
Maxwell's equations can be combined to give
\begin{equation}
c^2\epsilon_0\ppfrac{E}{x} 
= \ppfrac{P}{t} + \epsilon_0 \ppfrac{E}{t}
\label{Max_trans}
\end{equation}
Alternatively we assume that the electric and polarization fields
longitudinal, and the magnetic field vanishes. I.e. $\VE=E(t,x)\Ve_1$,
$\VP=P(t,x)\Ve_1$ and $\VB={\boldsymbol 0}$.
Maxwell's equations are
automatically satisfied if
\begin{equation}
\epsilon_0 E + P = 0
\label{Max_long}
\end{equation}

The permittivity relationship between $P(t,x)$ and $E(t,x)$ analysed
here is
\begin{eqnarray}
L(\omega, x) \FtP(\omega, x) +
\frac{\beta(\omega,x)^2}{(2\pi)^2}\ppfrac{\FtP}{x}(\omega, x)
=
\epsilon_0 \FtE(\omega, x)
\label{Intr_P_PDE}
\end{eqnarray}
where $\beta(\omega,x)$ is the finite propagation speed of the
polarisation wave multiplied by a constant, giving it dimensions of
length, and $L(\omega,x)$ is a dimensionless constitutive quantity. The
Fourier transform of $P(t,x)$ with respect to $t$ is given by
\begin{eqnarray}
\FtP(\omega,x) = \int_{-\infty}^{\infty} e^{-\twopii \omega t}
P(t,x) dt
\label{def_FFt}
\end{eqnarray}
If $L(\omega, x)=L(\omega)$ and $\beta(\omega, x)=\beta(\omega)$ are
independent of $x$ then we can Fourier transform (\ref{Intr_P_PDE}) to
give
\begin{eqnarray}
\FT{P}(\omega,k)
=
\frac{\epsilon_0 \FT{E}(\omega,k)}
{\displaystyle
L(\omega) -\beta(\omega)^2 k^2}
\label{Intro_homo_ft_CR}
\end{eqnarray}
where 
\begin{equation}
\FT{P}(\omega,k) = \int_{-\infty}^{\infty}
\int_{-\infty}^{\infty} e^{-\twopii(\omega t+ k x)} P(t,x) dt\, dx 
\end{equation}
is the Fourier transform of $P(t,x)$. Such a constitutive relation can
be made using diagonally crossed wires
\cite{costa2012macroscopic}. If
$L(\omega)={\big((\omega-i\lambda_0)^2+\omega_{\rm{P}}^2\big)}/
{\omega_{\rm{P}}^2}$
 then
(\ref{Intro_homo_ft_CR}) is the generalisation of a single-resonance
Lorentz model of permittivity \cite{lorentz1916theory} to include spatial
dispersion. For longitudinal modes, one can
also use a wire medium
\cite{belov2003strong,song2013accurate}.  Using
the longitudinal component of $\FT{\epsilon}$ given by
\cite[eqn. (4b)]{song2013accurate} \footnote{Comparing notation:
$2\pi \omega^{\textup{(here)}}=k^{\textup{(Song)}}$ and 
$2\pi k^{\textup{(here)}}=q_x{}^{\textup{(Song)}}$. We set
$a^{\textup{(Song)}}=b^{\textup{(Song)}}=b^{\textup{(here)}}$.}
\begin{equation}
\FT{P}(\omega,k) =
\frac{-\epsilon_0 \FT{E}(\omega,k)}
{\displaystyle
\omega^2/k_{\textup{p}}^2 -k^2/k_{\textup{p}}^2}
\label{Intro_Pol_Song}
\end{equation}
where, the grid spacing $b$ in the same in both $y$ and $z$ and
\begin{equation}
k_{\textup{p}}^2 
=
\frac{2\pi/b^2}{\log b - \log 2\pi r + 0.5275}
\label{Intro_Pol_Song_kp}
\end{equation}

For transverse modes, we consider a periodic structure where $L$ satisfies
\begin{eqnarray}
L(\omega,x+ a)=L(\omega,x)
\label{L_periodic}
\end{eqnarray}
and $\beta(\omega, x)=\beta(\omega)$ is independent of $x$.
We further assume that the amplitude of the inhomogeneity is
dominated by the first mode, that is
\begin{eqnarray}
L(\omega,x)=L_0(\omega) + 2\Lambda(\omega)\cos\left({{2\pi x}}/{a}\right)
\label{Per_L}
\end{eqnarray}
In the case of longitudinal modes, it is the plasma frequency
$k_{\textup{p}}$ given in (\ref{Intro_Pol_Song_kp}) which is periodic
in $x$, caused by a periodic variation in $r$.

In both the transverse and longitudinal cases, the goal in this
article is to find solutions to the source free Maxwell equations,
i.e. the dispersion relations. However since the medium is
inhomogeneous it is not possible to find single mode solutions of the
form $e^{\twopii (\omega t + k x)}$.  In section \ref{ch_Per} we
combine the polarisation differential equation (\ref{Intr_P_PDE}) with
Maxwell's equation to get a single 4th order ODE. Since this has
periodic coefficients, this is an example of Floquet's equation. Hence
we look for Floquet's modes of the form
\begin{equation}
P(t,x)
=
e^{\twopii\omega t}\FtP(x) 
= 
e^{\twopii\omega t}
\sum_{q=-\infty}^\infty P_{q} \, e^{2 \pi i (q + \kappa) x/a}
\label{Per_P_modes}
\end{equation}
where $\kappa$, $0\le\kappa<1$ is the phase. 
This gives rise to a difference equation for $P_q$
\begin{equation}
\Lambda(\omega) P_{q-1} + f_q(\omega) P_{q} + \Lambda(\omega) P_{q+1} = 0
\label{Per_diff_eqn}
\end{equation}
where in the transverse case
\begin{equation}
f_q(\omega) = L_0(\omega) - \frac{\beta(\omega)^2 (q + \kappa)^2}{a^2} + \frac{a^2\omega^2}{a^2\omega^2 - c^2 (q + \kappa)^2}
\label{Per_fk}
\end{equation}
This is shown in the appendix. In the Longitudinal
case, for a wire medium,
\begin{equation}
f_q(\omega)=\omega^2-k_0^2-\frac{(q+\kappa)^2}{a^2}
\label{wires_f_q}
\end{equation}
where $k_0^2=k_p^2|_{r=r_0}$ is a constant.

We can trivially solve
(\ref{Per_diff_eqn}) by arbitrarily fixing $P_0$ and $P_1$ and
then applying (\ref{Per_diff_eqn}) iteratively to find all subsequent
$P_q$. However, in general, this will lead to the coefficients $P_q$
diverging, i.e. 
$|P_{q}|\to\infty$ as $q\to\pm\infty$. The Fourier transform of this
would therefore be a non-smooth wave which may not even be
continuous. Therefore, for physical solutions, we demand that
\begin{eqnarray}
\label{pconverge}
|P_{q}|\to 0 \qquad \mbox{ as } \qquad q\to\pm\infty
\end{eqnarray}

\vspace{1em}

\begin{figure}
\setlength{\unitlength}{0.7cm}
\begin{picture}(11,9)(0,0)
\put(1,0){\includegraphics[viewport=27 8 544 548,
height=9\unitlength]{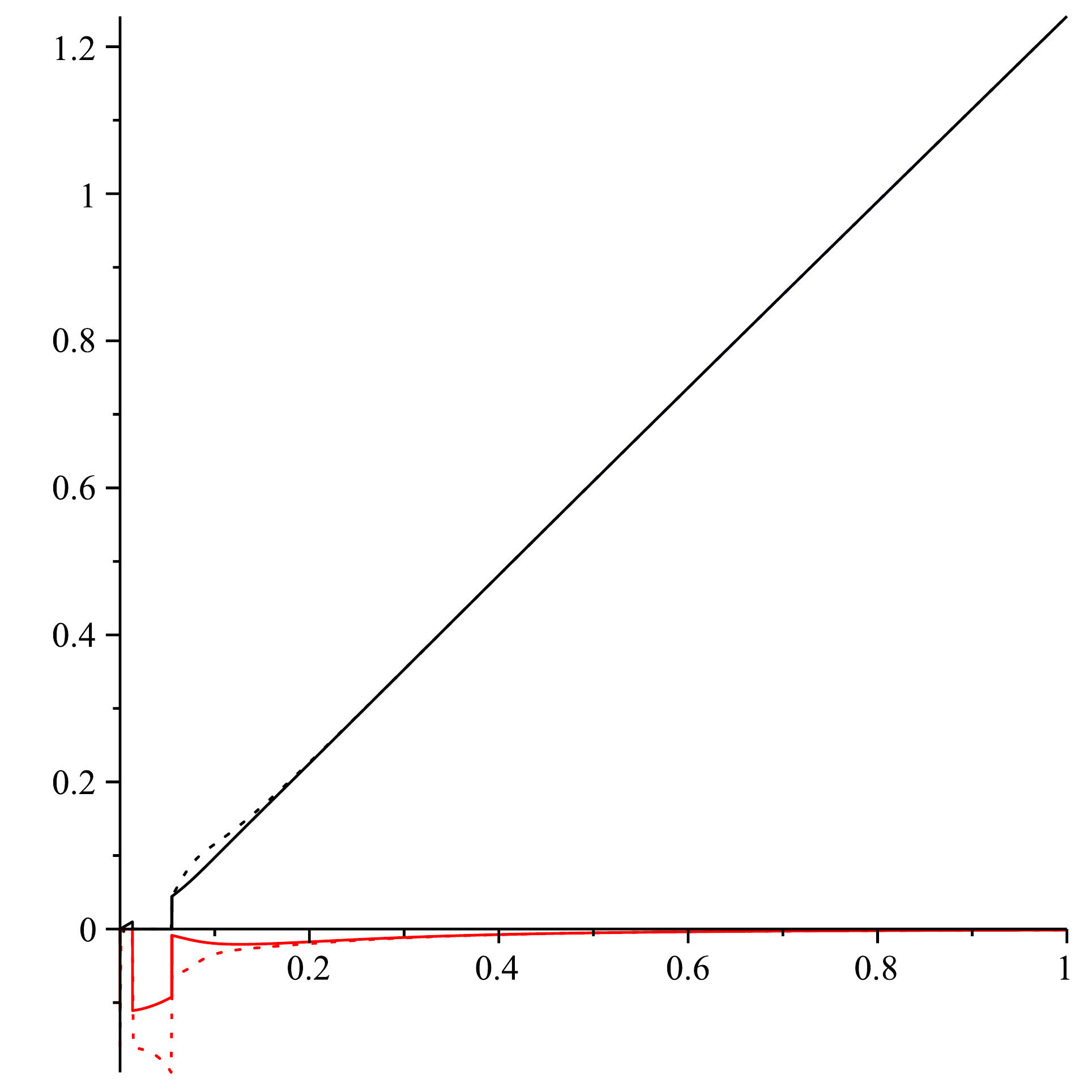}}
\put(0.,5){$\omega_0$}
\put(5.3,0){$\kappa$}
\end{picture}
\quad
\begin{picture}(11,9.2)(0,0)
\put(1,0.2){\includegraphics[viewport=27 8 544 548,
height=9\unitlength]{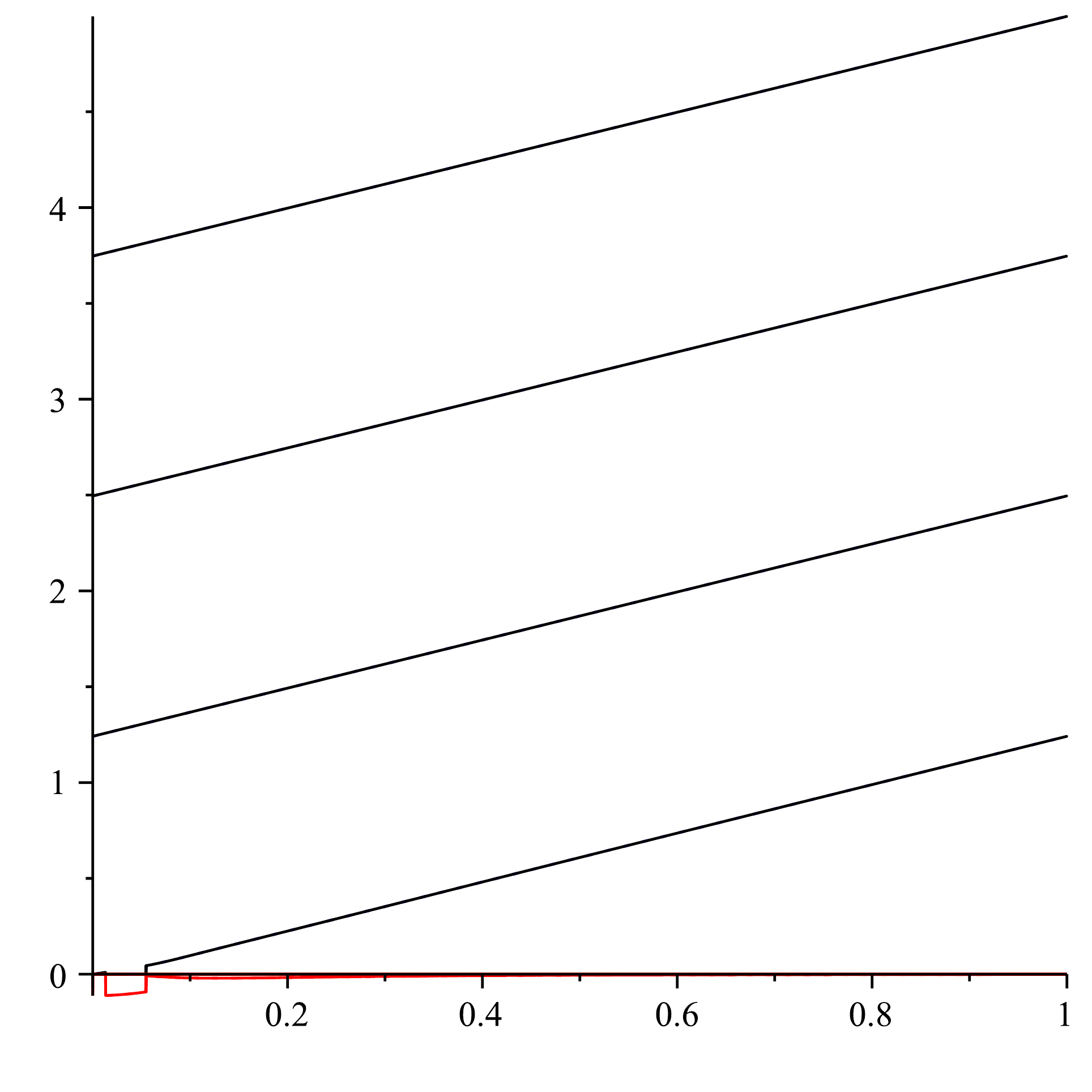}}
\put(0.,5){$\omega_n$}
\put(10.1,2.5){$\omega_0$}
\put(10.1,4.7){$\omega_1$}
\put(10.1,7.0){$\omega_2$}
\put(10.1,9.2){$\omega_3$}
\put(5.3,0){$\kappa$}
\end{picture}
\caption{The dispersion curve for the the transverse modes for single
  pole resonance $L(\omega) = {(\omega - i \lambda)^2}/{\omega_p^2} +
  1$ where $a = 0.8$, $\beta = 0.4$, $\lambda = 0.1$ and $\omega_p =
  0.15$. The lowest mode $n=0$ (left) and several modes (right) The
  curves are the unperturbed frequency $\Re(\Omega_0)$, (solid black),
  $\Im(\Omega_0)$ (solid red), and the perturbed frequency with
  $\Lambda=0.75$, $\Re(\omega_0)$ (dashed black), $\Im(\omega_0)$
  (dashed red).}
\label{fig_Dispersion_rel}
\end{figure}
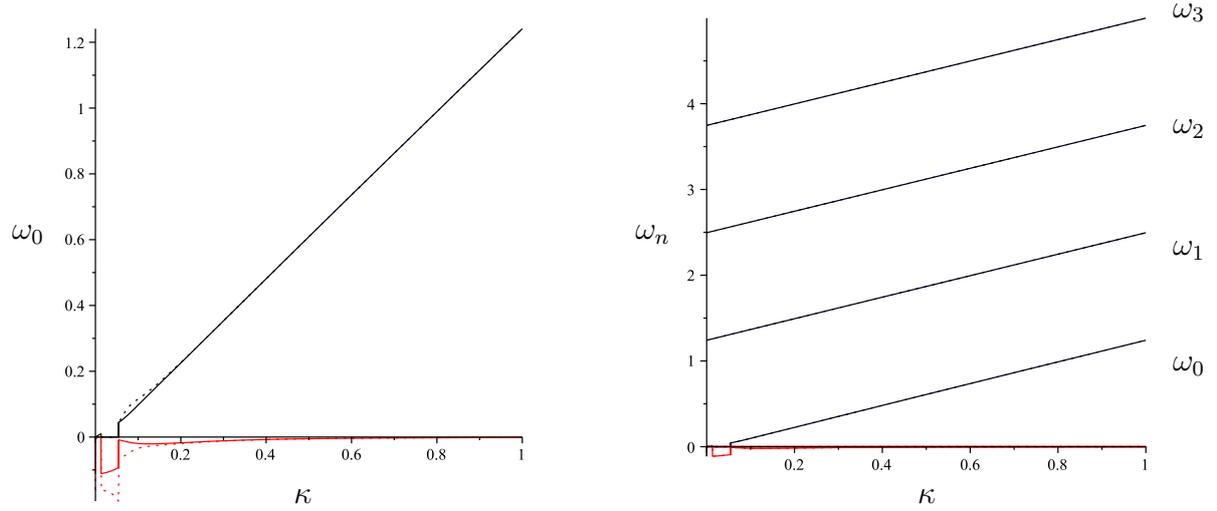

\begin{figure}
\setlength{\unitlength}{0.035\textwidth}
\begin{picture}(10.5,10.5)
\put(0.5,0.5){
\includegraphics[height=10\unitlength,
viewport=75 437 375 733]{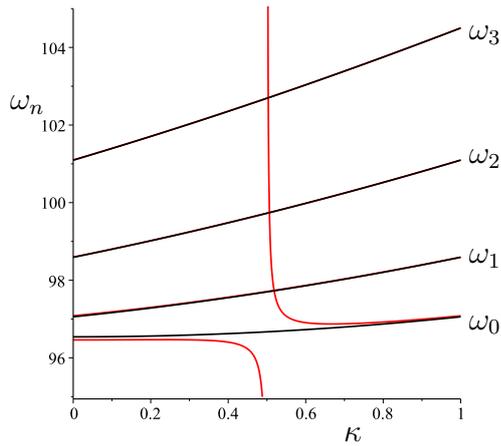}}
\put(0,8){$\omega_n$}
\put(8,0){$\kappa$}
\put(11,2.7){$\omega_0$}
\put(11,4.3){$\omega_1$}
\put(11,6.7){$\omega_2$}
\put(11,9.7){$\omega_3$}
\end{picture}
\caption{Dispersion curve for unperturbed (black) and perturbed (red)
  for longitudinal modes in a wire medium. Here $a = 0.1$, $b=0.02$, $r_0 =
  0.001$, and $r_1 = r_0/100$. Only the $n=0$ modes can we see the
  difference between the perturbed and unperturbed modes.}
\label{fig_long}
\end{figure}

In section \ref{ch_Approx} we make the assumptions that the
inhomogeneity $\Lambda(\omega)$ is a small constant.  This enables us to give a
dispersion relation for $\omega$ and $\kappa$, as well as the
harmonics $P_q$. In the unperturbed case $\Lambda=0$, then
(\ref{Per_diff_eqn}) becomes $f_q(\omega_n) P^n_q=0$, which we solve
by setting $\omega_n=\Omega_n$ and $P^n_q=\delta^n_q$ where
$f_n(\Omega_n)=0$.  We find that, analogous to non spatially
dispersive media, there exist an infinite set of modes,
$\big(\omega_n,\FtP^n(x)\big)$, where $n\in\Intg$. Thus we write
(\ref{Per_P_modes}) as
\begin{equation}
P^n(t,x)
=
e^{\twopii\omega_n t}\FtP^n(x) 
= 
e^{\twopii\omega_n t}
\sum_{q=-\infty}^\infty P^n_{q} \, e^{2 \pi i (q + \kappa) x/a}.
\label{Per_P_modes2}
\end{equation}
where
\begin{equation}
\omega_n=\Omega_n+O(\Lambda)
\label{Per_omega_n_pert}
\end{equation}
and
\begin{equation}
f_n(\Omega_n)=0
\label{Per_Omega_n}
\end{equation}
We ``normalise'' (\ref{Per_P_modes2}) by setting
\begin{equation}
P^n_n=1
\label{Per_P_modes_normalisation}
\end{equation}
The relationship between $\omega_n$ and $\kappa$ which satisfy
(\ref{Per_Omega_n}) corrected by the perturbation
(\ref{Per_omega_n_pert}) is a dispersion relation.  An
example of the dispersion relations, both of the unperturbed system and
the perturbed system, is given in figure
\ref{fig_Dispersion_rel} for the transverse case and figure
\ref{fig_long} for the longitudinal case.

We do not solve (\ref{Per_diff_eqn}) exactly. Instead we find approximate
solution to some order, that is we solve
\begin{equation}
\Lambda P^n_{q-1} + f_q(\omega_n) P^n_{q} + \Lambda P^n_{q+1} 
= 
O(\Lambda^r)
\label{Per_diff_eqn_On}
\end{equation}
for some order $O(\Lambda^r)$. In this article we solve
(\ref{Per_diff_eqn_On}) at least to order $O(\Lambda^3)$, which is
where the new features appear. However in many cases we can do much
better. For example, with the uncoupled modes given by (\ref{gsolendP})
below, $P^n_q=O(\Lambda^{|n-q|})$ so that we can trivially solve
(\ref{Per_diff_eqn_On}) to order $O(\Lambda^{|n-q|-1})$, although in
fact we solve (\ref{Per_diff_eqn_On}) to order
$O(\Lambda^{|n-q|+2})$.

There are two classes of solutions which we call \defn{uncoupled
  modes} and \defn{coupled modes}. 
The uncoupled modes have the property that
\begin{equation}
f_q(\Omega_n)\ne0
\qquad\mbox{for all}\quad q\ne n
\label{Per_f_q_generic}
\end{equation}
and are given by
\begin{equation}
\omega_n=\Omega_n+O(\Lambda)
\qquadand
P^n_q=\delta^n_q+O(\Lambda)
\label{Per_omega_P_generic}
\end{equation}
Solving \eqref{Per_diff_eqn},\eqref{Per_fk},\eqref{Per_omega_n_pert}
and \eqref{Per_Omega_n}, gives rise to a dispersion relation between
$\kappa$ and $\omega_n$. An example is given in figure
\ref{fig_Dispersion_rel}.

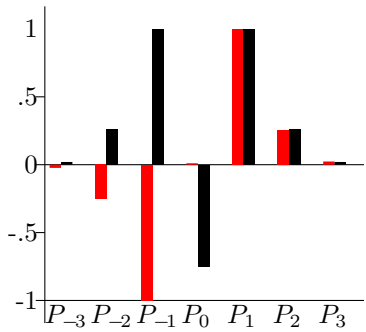
\begin{figure}
\setlength{\unitlength}{0.6cm}
\begin{picture}(8.5,7.5)(-3,0)
\put(0,0){\footnotesize $P_{-\!3}$}
\put(1,0){\footnotesize $P_{-\!2}$}
\put(2,0){\footnotesize $P_{-\!1}$}
\put(3,0){\footnotesize $P_{0}$}
\put(4,0){\footnotesize $P_{1}$}
\put(5,0){\footnotesize $P_{2}$}
\put(6,0){\footnotesize $P_{3}$}

\put(-0.2,0.5){\line(1,0){7.2}}
\put(0,0){\line(0,1){7}}
\put(-0.2,3.5){\line(1,0){7.2}}
\put(0,2){\line(-1,0){0.2}}
\put(0,5){\line(-1,0){0.2}}
\put(-0.15,0.5){\makebox(0,0)[r]{\footnotesize -1}}
\put(-0.15,2.0){\makebox(0,0)[r]{\footnotesize -.5}}
\put(-0.15,3.5){\makebox(0,0)[r]{\footnotesize 0}}
\put(-0.15,5.0){\makebox(0,0)[r]{\footnotesize .5}}
\put(-0.15,6.5){\makebox(0,0)[r]{\footnotesize 1}}

\put(0,0.5){
\scalebox{1}[3]{\color{red}
\begin{picture}(7,3)(-3,-1)
\linethickness{0.25\unitlength}
\put(-3,0){\line(0,-1){0.0236108}}
\put(-2,0){\line(0,-1){0.252416}}
\put(-1,0){\line(0,-1){1}}
\put( 0,0){\line(0,1){0.01}}
\put( 1,0){\line(0,1){1}}
\put( 2,0){\line(0,1){0.252416}}
\put( 3,0){\line(0,1){0.0236108}}
\end{picture}
}}
\put(0.25,0.5){
\scalebox{1}[3]{\color{black}
\begin{picture}(7,3)(-3,-1)
\linethickness{0.25\unitlength}
\put(-3,0){\line(0,1){0.02}}
\put(-2,0){\line(0,1){0.26}}
\put(-1,0){\line(0,1){1}}
\put( 0,0){\line(0,-1){0.75}}
\put( 1,0){\line(0,1){1}}
\put( 2,0){\line(0,1){0.26}}
\put( 3,0){\line(0,1){0.02}}
\end{picture}
}}
\end{picture}
\caption{The odd $\voddpnn{1}$ (red) and even $\vevenpnn{1}$ (black)
  transverse modes for $L_0\equiv 1$,
 $\Lambda\equiv 0.75$, $a \equiv 1$, $\beta \equiv 1$. In this case
 $\kappa = 0$, $\omodd_{1}=0.399$
 and $\omeven_{1}=0.753i$. }
\label{fig_P}
\end{figure}

\begin{figure}
\setlength{\unitlength}{0.035\textwidth}
\begin{picture}(10,11)
\put(1,0){\includegraphics[width=10\unitlength]{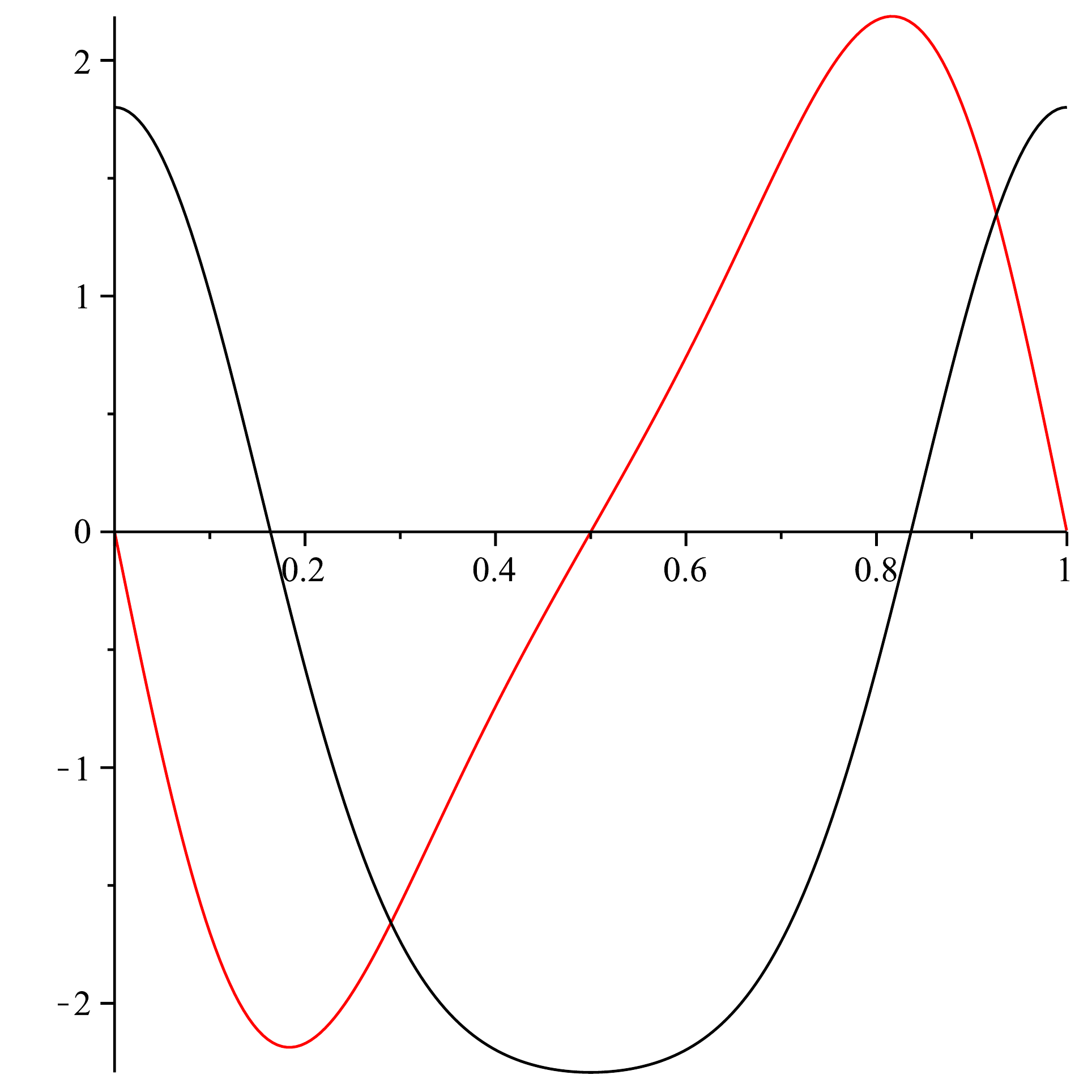}}
\put(11.,4.5){$x$}
\put(0,8){$\FtP(x)$}
\end{picture}
\caption{The odd $\hat{P}^\Evn_1(x)$ (black) and even
  $-i\hat{P}^\Odd_1(x)$ (red) transverse modes for $L_0$, $\Lambda$, $a$,
  $\beta$, $\omodd_{1}$ and $\omeven_{1}$ given in figure
  \ref{fig_P}. }
\label{fig_P1_x}
\end{figure}

The existence of coupled modes is a new feature that arises with spatially dispersive media. These occur when
(\ref{Per_f_q_generic}) is not satisfied. That is there exist integers
$n$ and $m$ and a frequency $\Omega_n\in\Cmpx$ 
such that $n\ne m$ and
\begin{equation}
f_n(\Omega_n)=f_m(\Omega_n)=0
\label{Per_coupled_f_nm}
\end{equation}
In this case the modes corresponding to $P^n$ and $P^m$,
(\ref{Per_P_modes2}), are coupled by the perturbation.
These are explored in section \ref{ch_Special}.  As stated we wish to
satisfy (\ref{Per_diff_eqn_On}) to order $O(\Lambda^3)$. This can
easily be achieved if $n-m\ge4$, where we say that coupled modes have
\defn{decoupled} to order $O(\Lambda^3)$.  However if $n-m\le3$ then
new solutions appear. These require taking higher order expansions for
$\omega_n$ and $P^n_q$. Furthermore if $n-m\le 2$ then the frequency
is shifted.  In the case when $\kappa=0$ one can use the symmetry in
order to construct the odd and even modes. An example of the lowest
two modes is given in figures \ref{fig_P} and \ref{fig_P1_x}.

\begin{figure}
\setlength{\unitlength}{0.035\textwidth}
\begin{picture}(11,10)
\put(1,0){
\includegraphics[height=10\unitlength,
viewport=76 424 375 716]{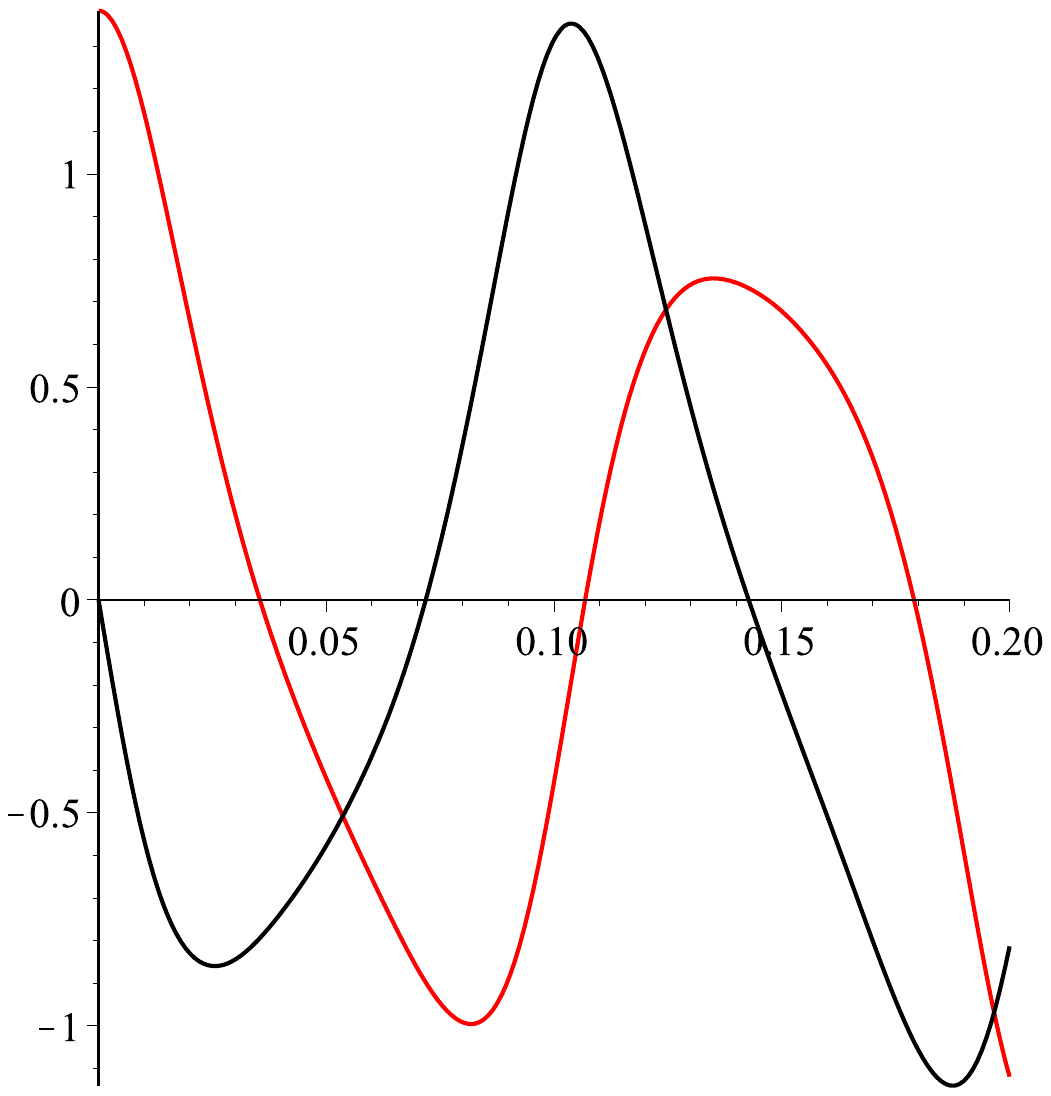}}
\put(0,8){$\FtP(x)$}
\put(11,3.5){$x$}
\end{picture}
\caption{Real (black) and Imaginary (red) longitudinal 
mode shape for $n=0$ and $\kappa=0.3$. Here the values of $a$, $b$ and
$r_0$ are given in figure \ref{fig_long} but $r_1=r_0/5$
 to see the distortion. Here 
 $P^{0}_{-2}=0.027$, $P^{0}_{-1}=0.174$,  $P^{0}_{0}=1$,
 $P^{0}_{1}=0.154$, $P^{0}_{2}=0.019$.} 
\label{fig_long_P}
\end{figure}
We see in figure \ref{fig_P1_x} that the spatial profile of the lowest
modes are significantly altered by the presence of the spatial dispersion. This
suggests that with higher order harmonics $\cos\left({4\pi x/a}\right)$ etc. in (\ref{Per_L}), one will be able to choose the
electric field profile.

In section \ref{ch_Numerical} we present an alternative 
numerical method for finding the permitted frequencies and
$\FTt{P}(x)$. 
Finally in the concluding section \ref{ch_Conc} we discuss the
application of inhomogeneous periodic structures.

\section{Floquet's Modes}
\label{ch_Per}

For transverse modes, taking the Fourier transform of
(\ref{Max_trans}) with respect
to $t$ 
gives
\begin{eqnarray}
(2\pi)^{-2}{\FtE''}=-\omega^2\big(c^{-2}\FtE+ \mu_{0} \FtP\big)
\label{Intr_Ej_FTt}
\end{eqnarray}
Here prime is the partial differentiation with respect to $x$.
In most cases, in the following, we will not explicitly
write the $\omega$ argument in $\FtE$, $\FtP$, $L$ and $\beta$.

Taking the Fourier transforms of (\ref{Intr_P_PDE}) and
(\ref{Intr_Ej_FTt}) with respect to $x$ we get 
\begin{eqnarray*}
&(\omega^2 c^{-2} -k^2)\tE(k) = -\omega^2 \mu_{0} \tP(k)
\end{eqnarray*}
and
\begin{eqnarray}
- k^2 \beta^2 \tP + (\tL *\tP)(k) = \epsilon_0 \tE(k)
\label{Per_FT_P_eqn}
\end{eqnarray}
Combining these into a single convolution equation gives
\begin{eqnarray}
  (\tL * \tP)(k) = \left(\beta^2 k^2 - \frac{
      \omega^2}{\omega^2 - k^2 c^2} \right) \tP(k)
\label{Per_Convolution_eq}
\end{eqnarray}
or equivalently
\begin{eqnarray}
\frac{\beta^2}{(2\pi)^2}
\frac{\partial^4 \hat{P}(x)}{\partial x^4} 
&+ \left(L(x) + \frac{\beta^2 \omega^2}{c^2} \right) \frac{\partial^2 \hat{P}(x)}{\partial x^2} 
+ 2\frac{\partial L(x)}{\partial x} 
\frac{\partial \hat{P}(x)}{\partial x}\nonumber\\
&+ \left(\frac{\partial^2 L(x)}{\partial x^2} 
+ \frac{2 \pi \omega^2}{c^2}L(x)P(x) +\frac{2\pi \omega^2}{c^2} \right)\hat{P}(x) 
= 0
\label{Per_4th_ODE}
\end{eqnarray}
in $(t,x)$ space.
We see that (\ref{Per_4th_ODE}) is a version of Floquet's
equation. Hence $\FtP(x)$ is given by 
\begin{equation}
\FtP(x) = \sum_{q=-\infty}^\infty P_{q} \, e^{2 \pi i (q + \kappa) x/a}
\label{Per_P_anzats}
\end{equation}
hence (\ref{Per_P_modes}).
In the appendix we show that from (\ref{Per_L}),
\eqref{Per_Convolution_eq} and (\ref{Per_P_anzats}) the coefficients
$P_q$ satisfy the difference equation \eqref{Per_diff_eqn} with $f_q$
given by \eqref{Per_fk}.

Observe that having higher harmonics, such as 
$\cos\left({4{\pi}x/a}\right)$ in  (\ref{Per_L}), 
will result in (\ref{Per_diff_eqn}) being replaced by a higher order
difference equation. Likewise letting $\beta$ depend periodically on
$x$ will also increase the order of the difference equation.

\subsection{Longitudinal modes in wire media with 
periodic variation}
\label{ch_wires}
\begin{figure}
\includegraphics[height=3cm]{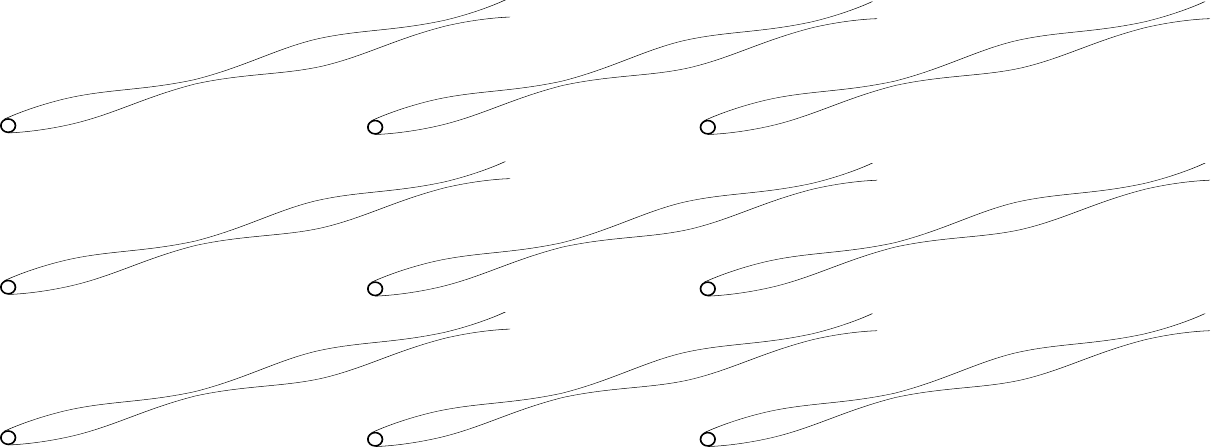}
\caption{Wire medium with periodically varying wires}
\label{fig_wires}
\end{figure}

The case for longitudinal modes is very similar. Using
\eqref{Intro_Pol_Song},\eqref{Intro_Pol_Song_kp} with the radius of the
wire varying periodically, see figure \ref{fig_wires}
\begin{equation}
r(x)=\frac{b}{2\pi}
\exp\Big(\frac{-2\pi/b^2}{k_0^2+2\Lambda\cos(2\pi x/a)} + 0.5275\Big)
\label{wires_r_x}
\end{equation}
where $k_0^2=k_{\textup{p}}^2|_{r=r_0}$. 
Substituting into (\ref{Intro_Pol_Song_kp}) we see
\begin{equation}
k_{\textup{p}}^2=k_{0}^2 + 2\Lambda\cos(2\pi x/a)
\end{equation}
We may consider simply setting $r(x)=r_0+r_1\cos(2\pi x/a)$, which
will produce terms such as $\cos(2\pi x/a)^2$ and will increase the
degree of the difference equation (\ref{Per_diff_eqn}). For small
enough $r_1$ then we can ignore the higher terms $\cos(2\pi x/a)^2$,
then 
$\Lambda\approx 
(r_1 k_{\textup{p}}^4 b^2)/(2\pi r_0)$.
Combining
(\ref{Per_FT_P_eqn}), (\ref{Max_long}) and (\ref{Per_P_anzats}) we
once again get the difference equation (\ref{Per_diff_eqn}) where
$f_q(\omega)$ is given by (\ref{wires_f_q}).
The dispersion relation for $\omega$ and $\kappa$ is plotted in figure
\ref{fig_long}. Observe the the perturbed modes are dramatically
different near $n=0$ and $\kappa=\frac12$. This is near the point
where the modes couple. The detailed analysis of this behaviour is
future work. An example distorted mode is given in figure
\ref{fig_long_P}.  The result is a three dimensional periodic
structure, with periods $(a,b,b)$. This could also be analysed using
numerical techniques, however this would be more time consuming.

\section{Approximate Analytic solutions}
\label{ch_Approx}

As stated, we assume here that the inhomogeneity of the medium given
by $\Lambda$ in (\ref{Per_L}) is small and give approximate analytic
solutions to \eqref{Per_diff_eqn} which are valid to at least
$O(\Lambda^3)$.

Let $\Omega_n$ be a solution to (\ref{Per_Omega_n}), 
$f_n(\Omega_n)=0$.
For each $\Omega_n$ we can find the corresponding $P^n_q$ such that
(\ref{Per_diff_eqn}) is satisfied to some order given below. 

In the case of transverse modes, from
(\ref{Per_fk}) it is clear that $f_{n}(\omega)=F(Q,\omega)$ where
\begin{equation}
F(Q,\omega) = L_0(\omega) - \frac{\beta(\omega)^2 Q^2}{a^2} + \frac{a^2\omega^2}{a^2\omega^2 - c^2  Q^2}
\label{Approx_F_trans}
\end{equation}
where $Q=n+\kappa$ and hence $\kappa=Q(\rm{\,mod\,}1)$ and $n=\lfloor
Q\rfloor$, the largest integer $n\le Q$. Thus giving the shape of
figure \ref{fig_Dispersion_rel}. The perturbation given in
(\ref{gsol1}) below also has this property.
In the longitudinal case $F(Q,\omega)$ is
replaced by $F(Q,\omega)=\omega^2-k_0^2-Q^2/a^2$.

In this section we consider the generic case, i.e. that there are no
other integer $m$ such that $f_m(\Omega_n)=0$. The special cases when
there exists $n\ne m\in\Intg$ and $\Omega_n$ such that
$f_n(\Omega_n)=f_m(\Omega_n)=0$ are considered below in section
\ref{ch_Special}.
Let
\begin{eqnarray}
  \omega_n &= 
\omo_{n} + 
\Lambda^2 
\Big(\frac{1}{\FOmega'_n\FOmega_{n-1}} + \frac{1}{\FOmega'_n\FOmega_{n+1}}\Big) + O(\Lambda^4) 
\label{gsol1}
\end{eqnarray}
and
\begin{eqnarray}
P^n_q =
\left\{
\begin{array}{ll}
\displaystyle
\frac{(-\Lambda)^{|q-n|}}{\prod_{k=1}^{|q-n|}\FOmega_{n+k}} +
O(\Lambda^{|q-n| + 2}) &\qquad q > n
\\[2em]
\displaystyle
1 &\qquad q = n
\\
\displaystyle
\frac{(-\Lambda)^{|q-n|}}{\prod_{k=1}^{|q-n|}\FOmega_{n-k}} 
+ O(\Lambda^{|q-n| + 2}) &\qquad q<n
\end{array}
\right.
\label{gsolendP}
\end{eqnarray}
where
\begin{equation}
\FOmega_q=f_{q}(\Omega_n)
\,,\qquad
\FOmega'_q=\frac{d\,f_{q}}{d\omega}\Big|_{\Omega_n}
\quad\mbox{and}\quad
\FOmega''_q=\frac{d^2\,f_{q}}{d\omega^2}\Big|_{\Omega_n}\label{def_F_Omega}
\end{equation}
These $(\omega_n,P^n_q)$ solve the
difference equation \eqref{Per_diff_eqn} to an order of $\Lambda$
which depends on $n$ and $q$ as follows
\begin{eqnarray}
&\Lambda P^n_{q-1} + f_{q}(\omega_n) P^n_q + \Lambda P^n_{q+1} =
O(\Lambda^{|q-n| +2})  
\qquad\mbox{for\ \  $q \neq n$ }
\label{thmg1}
\end{eqnarray}
and
\begin{eqnarray}
\Lambda P_n^{n-1} + f_{n}(\omega_n) P_n^{n} + \Lambda P_n^{n+1} =
O(\Lambda^{4})  
\label{thmg2}
\end{eqnarray}
The proof of (\ref{thmg1}) and (\ref{thmg2}) is given in the
appendix. As in standard perturbation theory, improvement in the
accuracy of (\ref{gsol1}) and (\ref{gsolendP}) can be made by higher
order corrections.

\section{Coupled Modes}
\label{ch_Special}

In this section we consider the case when there are two distinct integers
$n$ and $m$ 
and $\Omega$  such that
$f_{n}(\Omega)=f_{m}(\Omega)=0$.
By writing (\ref{Per_fk}) as a quadratic for $(q+\kappa)^2$ we see
that 
\begin{equation}
(q+\kappa)^2 = C_+(\Omega)^2 
\qquadtext{or}
(q+\kappa)^2 = C_-(\Omega)^2 
\label{Per_q+kappa}
\end{equation}
where
\begin{equation}
C_\pm(\Omega)^2 = \frac{a^2}{2}\left({\frac{\Omega^2}{c^2} +
    \frac{L_{0}}{\beta^2} \pm \sqrt{\left(\frac{\Omega^2}{c^2} +
        \frac{L_{0}}{\beta^2} \right)^2 - \frac{4\Omega^2(L_0 +
        1)}{\beta^2 c^2}}}\right)
\label{Per_q_Cpm}
\end{equation}
One of two situation can occur.
First, $n$ and $m$ correspond to the same root $C_\pm(\Omega)$, so
that
\begin{equation}
(n+\kappa)^2=(m+\kappa)^2
\label{Special_n_m_simple}
\end{equation}
This implies $2\kappa=-n-m$ and
so either $\kappa=0$ or $\kappa=\frac12$. These will be considered in
sections \ref{Sec_kappa0} and \ref{Sec_kappa_half} below.

In the second case set
\begin{equation}
(n+\kappa)^2 = C_+(\Omega)^2 
\qquadtext{and}
(m+\kappa)^2 = C_-(\Omega)^2 
\label{Per_m+kappa}
\end{equation}
Then eliminating $\kappa$ we have either
\begin{equation}
\big(C_+(\Omega)-C_-(\Omega)\big)\in\Intg
\qquadtext{or}
\big(C_+(\Omega)+C_-(\Omega)\big)\in\Intg
\label{Per_C_pm_Z}
\end{equation}
Since, for most of $\omega$, $C_+(\omega)\pm C_-(\omega)$ are
continuous function these situations will also occur in general.

In the Longitudinal case, from (\ref{wires_f_q}) we see that only
the case (\ref{Special_n_m_simple}) can occur.

Since two modes $(\omega_n,P^n_q)$ and $(\omega_m,P^m_q)$ are coupled,
we find that, in all cases, there are two new modes.

It may be possible for specially chosen $L_0(\omega)$ and
$\beta(\omega)$ that there may exist three or four roots integer roots
to (\ref{Per_fk}). However these will not be considered. Similarly one
can construct $L_0(\omega)$ and $\beta(\omega)$ such that
$f'_n(\Omega)=0$ or $f'_m(\Omega)=0$. In which case formulae such as
(\ref{case1_omega}) will not be valid. These will also not be considered.

In the special cases the dispersion curves become isolated points
relating $\omega$ and $\kappa$. For example when $\kappa=0$ and 
$f(\Omega)=0$ .

Recall our goal is to solve
(\ref{Per_diff_eqn_On}) to at least order $O(\Lambda^3)$. We define
the right hand side of (\ref{Per_diff_eqn_On}) as
$\mathcal{Q}^{n}_{q}$, i.e.
\begin{equation}
  \mathcal{Q}^{n}_{q} 
= \Lambda P^{n}_{q-1} + f_{q}(\omega_{n}) P^{n}_{q} + \Lambda P^{n}_{q+1}
\label{Per_def_Qn}
\end{equation}
so that we require $\mathcal{Q}^{n}_{q}=O(\Lambda^3)$. 

\subsection{Decoupled coupled modes.}
\label{Sec_other}

In the case when $n-m\ge 4$ then the coupled modes, decouple to order
$O(\Lambda^{3})$, then the two solutions have the same $\omega_n$
given by (\ref{gsol1}) and (\ref{Per_Omega_n}). The corresponding
modes are given by
\begin{eqnarray}
P^n_q = 
\left\{
\begin{array}{ll}
\displaystyle\frac{(-\Lambda)^{q-n}}{\prod_{k=1}^{q-n}\F_{n+k}} + O(\Lambda^{q-n + 2}) &\qquad q > n
\\
1 &\qquad q=n
\\
\displaystyle\frac{(-\Lambda)^{n-q}}{\prod_{k=1}^{n-q}\F_{n-k}} + O(\Lambda^{n-q + 2}) &\qquad m < q < n
\\
O(\Lambda^{n-m}) &\qquad q = m
\\
O(\Lambda^{n-q-2}) &\qquad q < m
\end{array}
\right.
\label{sol_other_n}
\end{eqnarray}
This satisfies (\ref{Per_def_Qn}), with $\mathcal{Q}^{n}_{n} =
O(\Lambda^{4})$, $\mathcal{Q}^{n}_{m} = O(\Lambda^{n-m})$, 
$\mathcal{Q}^{n}_{m-1} = O(\Lambda^{n-m+1})$, 
$\mathcal{Q}^{n}_{q} = O(\Lambda^{q - n + 2})$ for $q > m, q \neq n $
and
$\mathcal{Q}^{n}_{q} = O(\Lambda^{n-q-2})$ for $q < m-1$. The proof is
by direct substitution.
Observe that in \eqref{sol_other_n} there are terms that are given
only to $O(\Lambda^r)$. These terms are always at least order
$O(\Lambda^3)$ and furthermore it is clear that 
\eqref{Per_diff_eqn_On} is satisfied to at
least order $O(\Lambda^3)$. 

The second solution, $P^m_q$, is given by
\begin{equation}
P^m_q = P^n_{q'} \qquad \qquad \qquad \mbox{where}\qquad q' = n + m - q
\label{sol_other_m}
\end{equation}

\subsection{Coupled modes,  $n-m=1$}

In this case $n-m=1$ we have two solutions given by
\begin{eqnarray}
\fl
\omega_n = \Omega_n + \Lambda \omega'_n + 
\Lambda^2 \left(
\frac{1}{2\FOmega_{n-2}\FOmega'_{n-1}}+
\frac{1}{2\FOmega_{n+1}\FOmega'_{n}}-
\frac{1}{4 \FOmega'_{n-1}\FOmega'_{n}}
\Big(\frac{\FOmega''_{n}}{\FOmega'_{n}}+\frac{\FOmega''_{n-1}}{\FOmega'_{n-1}}\Big)
\right)
+O(\Lambda^3)
\nonumber
\\
\label{case1_omega}
\end{eqnarray}
where
\begin{equation*}
\omega'_n=\pm (\FOmega'_n\, \FOmega'_{n-1})^{-1/2}
\end{equation*}
The corresponding modes are given by
\begin{eqnarray}
\fl
P_{n-3} 
&=&
-\Lambda^2\frac{\FOmega'_n\omega'_n }{\FOmega_{n-2}\FOmega_{n-3}}+ O(\Lambda^3)
\,,
\nonumber
\\
\fl
P_{n-2} 
&=&  
\Lambda\frac{\FOmega'_{n}\omega'_n}{\FOmega_{n-2}} - 
\frac{\Lambda^2}{\FOmega'_{n-1}\FOmega_{n-2}}
\left( 
\frac{\FOmega''_{n-1}}{4\FOmega'_{n-1}}-
\frac{\FOmega''_{n}}{4\FOmega'_{n}}+
\frac{\FOmega'_{n-1}}{2\FOmega_{n+1}}-
\frac{\FOmega'_{n}}{2\FOmega_{n-2}}+
\frac{\FOmega'_{n-2}}{\FOmega_{n-2}}
\right)
+O(\Lambda^3)
\,,
\nonumber
\\
\fl
P_{n-1} 
&=& 
-\FOmega_n'\omega'_n + 
\frac{\Lambda}{\FOmega'_{n-1}}\left( 
\frac{\FOmega''_{n-1}}{4\FOmega'_{n-1}}-
\frac{\FOmega''_{n}}{4\FOmega'_{n}}+
\frac{\FOmega'_{n-1}}{2\FOmega_{n+1}}-
\frac{\FOmega'_{n}}{2\FOmega_{n-2}}\right)
+O(\Lambda^2)
\,,\qquad
P_1=1
\,,
\nonumber
\\
\fl
P_{n+1} 
&=& 
- \frac{\Lambda}{\FOmega_{n+1}} + \frac{\Lambda^2 \FOmega'_{n+1}\omega'_n}{(\FOmega_{n+1})^2} + O(\Lambda^3)
\,,\qquad
P_{n+2} = \frac{\Lambda^2}{\FOmega_{n+2}\FOmega_{n+1}} + O(\Lambda^3)
\label{case1_P}
\end{eqnarray}
Equations (\ref{case1_omega}) and (\ref{case1_P}) satisfy
(\ref{Per_diff_eqn_On}) to order $O(\Lambda^3)$. This follow from direct
substitution.

\subsection{Coupled modes,   $n-m=2$}

Of all the cases the case when $n-m=2$ is the most complicated. The
two modes are coupled at order $O(\Lambda^2)$. Therefore it is
necessary to take the second order perturbation to see the
coupling. The solution for $\omega_n$ is given by
\begin{eqnarray}
\omega_n 
&=& \Omega_n + 
\Lambda^2
\Big(\frac{\alpha+1}{\FOmega'_{n}\FOmega_{n-1}}
+\frac{1}{\FOmega'_{n}\FOmega_{n+1}}\Big) 
+ O(\Lambda^3)
\label{case2_omega}
\end{eqnarray}
where $\alpha$ is the solution to the quadratic
\begin{equation}
{\FOmega'_{n-2}}{} \alpha^2
+
\Big(\frac{\FOmega_{n-1} \FOmega'_{n-2}}{\FOmega_{n+1}} + {\FOmega'_{n-2}}{} -
\frac{\FOmega_{n-1} \FOmega'_n}{\FOmega_{n-3}} - {\FOmega'_n}{}\Big)\alpha
- {\FOmega'_n}{}
=0
\label{case2_alpha}
\end{equation}
The mode is described by
\begin{eqnarray}
P_{n-4} 
= \Lambda^2 \frac{\alpha}{\FOmega_{n-3} \FOmega_{n-4}} + O(\Lambda^3)
\,,\qquad
P_{n-3} 
= - \Lambda \frac{\alpha}{\FOmega_{n-3}} - \Lambda^2 \frac{\alpha_2}{\FOmega_{n-3}} + O(\Lambda^3)
\,,\nonumber\\
P_{n-2} 
=  \alpha + \Lambda \alpha_2 + O(\Lambda^2)
\,,\qquad
P_{n-1} 
= -\Lambda \frac{ \alpha + 1}{\FOmega_{n-1}} 
- \Lambda^2 \frac{\alpha_2}{\FOmega_{n-1}} + O(\Lambda^3)
\,,\nonumber\\
P_n 
= 1
\,,\quad
P_{n+1} 
= - \frac{\Lambda}{\FOmega_{n+1}} + O(\Lambda^3)
\,,\quad
P_{n+2} 
= \frac{\Lambda^2}{\FOmega_{n+2}\FOmega_{n+1}} + O(\Lambda^3)
\label{case2_P}
\end{eqnarray}
We see that the $\Lambda$ coefficients of $P_{n-2}$ and the 
$\Lambda^2$ coefficients of $P_{n-3}$ and $P_{n-1}$ are related by the
quantity $\alpha_2$. However it would require perturbing to order $O(\Lambda^4)$ to determine this quantity.

\subsection{Coupled modes,  $n-m=3$}
In the case $n-m=3$ one solution is given by
\begin{eqnarray}
\omega_n &= \Omega_n + \frac{\Lambda^2 }{\FOmega'_n} 
\left( \frac{1}{\FOmega_{n-1}} - \frac{1}{\FOmega_{n+1}} \right) + O(\Lambda^3)
\label{case3_omega}
\end{eqnarray}
and the mode given by
\begin{eqnarray}
P_{n-4} = - \Lambda^2 \frac{\alpha_3}{\FOmega_{n-4}}
\,,\quad
P_{n-3} = \Lambda \alpha_3 + O(\Lambda^2)
\,,
\nonumber\\
P_{n-2} =  \Lambda^2 \left( \frac{1}{\FOmega_{n-1} \FOmega_{n-2}} - \frac{\alpha_3}{\FOmega_{n-2}} \right) + O(\Lambda^3)
\,,\quad
P_{n-1} = -\frac{\Lambda}{\FOmega_{n-1}} + O(\Lambda^3)
\,,
\nonumber\\
P_n = 1
\,,\quad
P_{n+1} = - \frac{\Lambda}{\FOmega_{n+1}} + O(\Lambda^3)
\,,\quad
P_{n+2} = \frac{\Lambda^2}{\FOmega_{n+1} \FOmega_{n+2}} + O(\Lambda^3)
\label{case3_P}
\end{eqnarray}
where $\alpha_3$ cannot be determined order at order $O(\Lambda^3)$.

The other solution is given by exchanging $n$ and $m$, i.e. by
(\ref{case3_omega}) and (\ref{case3_P}) where $m-n=3$.

\subsection{Coupled modes when $\kappa = 0$}
\label{Sec_kappa0}

When $\kappa=0$ then from (\ref{Per_fk}) we see that
$f_q(\omega)=f_{-q}(\omega)$ for all $\omega$. Hence
$\FOmega_q=\FOmega_{-q}$ and $\FOmega'_q=\FOmega'_{-q}$.
Thus since $f_{n}(\Omega_n) = 0$ then $f_{-n}(\Omega_n) =0$. 
If $n\ne1,-1$ then we have the decoupled mode, with solution given by
(\ref{sol_other_n}) and (\ref{sol_other_m}).

The remaining case is when $n=1$. In this case some simplification
arises and the solutions may be interpreted as odd and even modes.
The frequencies are
\begin{eqnarray}
  &\omodd_{1} = \omo_{1} + 
  \frac{\Lambda^2}{\FOmega'_1\FOmega_2}
+ O(\Lambda^4)
\label{kappa0_omega_odd}
  \\
  &\omeven_{1} = \omo_{1} + \frac{\Lambda^2}{\FOmega'_1}
  \Big(\frac{1}{\FOmega_2} + \frac{2}{\FOmega_0} \Big) +
  O(\Lambda^4)
\label{kappa0_omega_even}
\end{eqnarray}
and the corresponding $P^{n}_{q}$
\begin{eqnarray}
P^{1, \Evn}_{q} = 
\left\{
\begin{array}{ll}
\displaystyle
R^1_q &\qquad q > 0
\\
\displaystyle
{-2 \Lambda}/{\FOmega_0} &\qquad q = 0
\\
\displaystyle
R^1_{-q} &\qquad q < 0
\end{array}\right.\label{solendlow}
\end{eqnarray}
and
\begin{eqnarray}
P^{1, \Odd}_{q} = 
\left\{
\begin{array}{ll}
\displaystyle
R^1_q &\qquad q > 0
\\
\displaystyle
0 &\qquad q = 0
\\
\displaystyle
-R^1_{-q} &\qquad q < 0
\end{array}\right.\label{solendlowodd}
\end{eqnarray}
where $R^1_1=1$ and
\begin{eqnarray*}
R^1_q =
\frac{(-\Lambda)^{q-1}}{\prod_{k=1}^{q-1}\FOmega_{1+k}} +
O(\Lambda^{q+1}) \quad\mbox{for}\quad q > 1
\end{eqnarray*}
These satisfy $\mathcal{Q}^{1}_{\pm1} = O(\Lambda^4)$,
$\mathcal{Q}^{1}_{0} = O(\Lambda^3)$ and $\mathcal{Q}^{1}_{q} =
O(\Lambda^{|q|+1})$ for $|q|\ge 2$ and $\mathcal{Q}^{1}_{1} =
O(\Lambda^4)$. The proof following similar to (\ref{thmg1}) and
(\ref{thmg2}) given in the appendix. This is consistent with
(\ref{case2_omega}) and (\ref{case2_P}) where from (\ref{case2_alpha})
$\alpha=\pm1$.

\subsection{Case $\kappa = \frac12$}
\label{Sec_kappa_half}

When $\kappa=\frac12$ then from (\ref{Per_fk}) we see that
$f_q(\omega)=f_{-1-q}(\omega)$ for all $\omega$. Hence
$\FOmega_q=\FOmega_{-1-q}$ and $\FOmega'_q=\FOmega'_{-1-q}$.
Thus since $f_{n}(\Omega_n) = 0$ then $f_{-1-n}(\Omega_n) =0$. 

Again if $n\ge3$ then the modes decouple. However if $n=0$ or $n=1$
then the modes remain coupled and we can use the results of $n-m=1$ or
$n-m=3$ respectively.
Unlike the case when $\kappa=0$ no significant simplification takes
place. However we note that (\ref{case1_omega}) simplifies to
\begin{eqnarray}
\fl
\omega_n = \Omega_n \pm \frac{\Lambda}{\FOmega'_0} - 
\Lambda^2 \left(
\frac{1}{\FOmega_{1}\FOmega'_{0}}-
\frac{\FOmega''_{0}}{2 \FOmega'_{-1}(\FOmega'_{0}{})^3}
\right)
+O(\Lambda^3)
\nonumber
\\
\label{kappa12_omega}
\end{eqnarray}


\section{Numerical Approaches} 
\label{ch_Numerical}

A numerical
approximation scheme, which is valid if $\Lambda\not\ll L_0$,
is as follows: Choose an integer $N\ge 2$. Then assume that
$P_m\approx 0$ for $|m|>N$ thus truncating the infinite set of
equation given by (\ref{Per_diff_eqn}) to a set of $2N+1$ linear
equations for $\Set{P_{-N},\ldots,P_{N}}$. Write this in matrix
language $M \underline b = \underline 0$ where $M$ is a
$(2N+1)\times(2N+1)$ matrix with $M_{k,k}=f_{k-N-1}$,
$M_{k,k-1}=M_{k,k+1}=\Lambda$ and $\underline b_k=P_{k-N-1}$.  Solve
$\det(M)=0$ to obtain values for $\omega$. The corresponding
null spaces give $P_m$.

For example, for $N = 2$ the matrix equation is
\begin{eqnarray*}
\left(
\begin{array}{ccccc}f(-2) & \Lambda & 0 & 0 & 0 \\ \Lambda & f(-1) &
  \Lambda & 0 & 0 \\ 0 & \Lambda & f(0) & \Lambda & 0 \\ 0 & 0 &
  \Lambda & f(1) & \Lambda \\ 0 & 0 & 0 & \Lambda & f(2)  \end{array}
\right) \left(\begin{array}{c} P_{-2} \\ P_{-1} \\ P_{0} \\ P_{1} \\
    P_{2} \end{array} \right) = \left(\begin{array}{c} 0 \\ 0 \\ 0 \\
    0 \\ 0 \end{array} 
\right)
\end{eqnarray*}

\section{Conclusion}
\label{ch_Conc}
In this article we describe a model for spatially dispersive
media with a periodic structure. Two types of media are considered,
one based on a single-resonance Lorentz
model of permittivity, the other on a wire medium.
Approximate analytic solutions to
Maxwell's equations are found for such media for small magnitudes in the
inhomogeneities. We demonstrate the existence of coupled modes.
Higher order correction should be possible by
repeating the procedure. If one were to solve (\ref{Per_diff_eqn_On})
to order $O(\Lambda^4)$ then one would expect coupled modes where
$n-m=4$ would remain coupled.

By combining the solutions above with additional boundary conditions 
\cite{
pekar1958theory,
henneberger1998additional,
maslovski2010generalized,
silveirinha2006additional,
silveirinha2008additional,
silveirinha2009additional,
gratus2013Inhomogeneous}
one will be able to predict the behaviour or electromagnetic radiation
as it passes though a slab of periodic spatially dispersive medium.

It should be possible to construct a meta-material
with such periodic spatially dispersive properties, especially the
wire medium. These can be used to test that the packets given above
are indeed supported. In addition, numerical simulations using, for
example, FDTD codes in a periodic cell, can be made.

\begin{figure}
\setlength{\unitlength}{0.03\textwidth}
\begin{picture}(20,12)
\put(0,0){\includegraphics[height=10\unitlength]{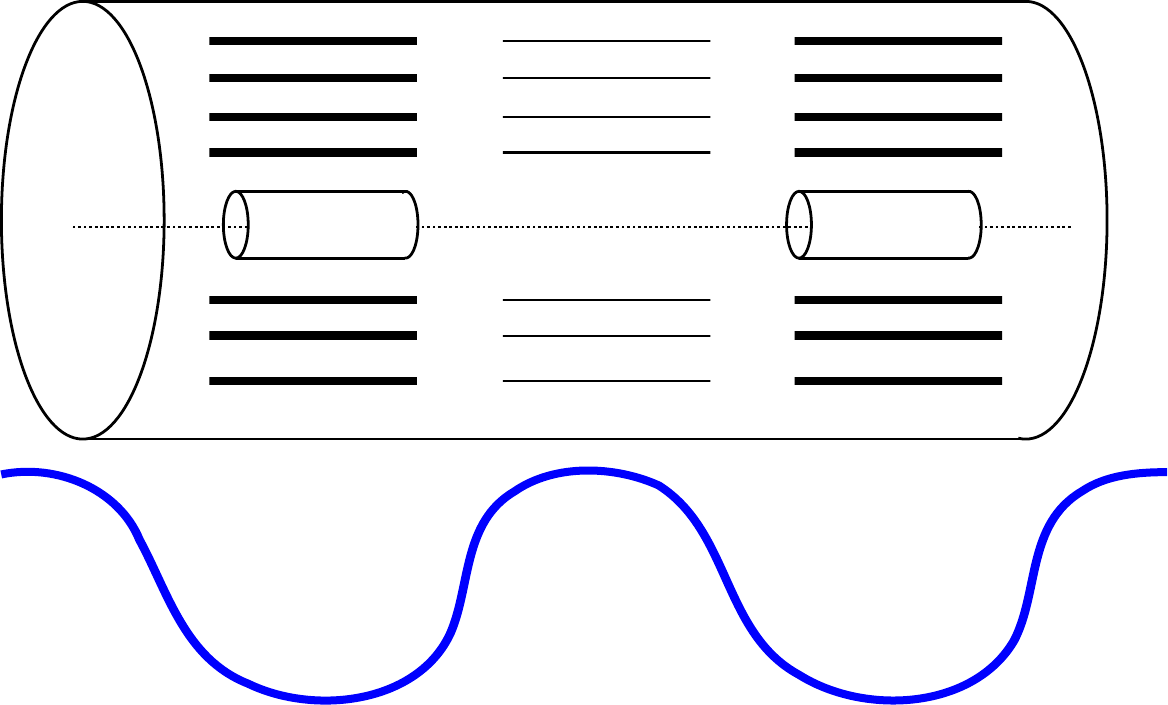}}
\put(18,9){\vector(-4,-1){3}}
\put(18.5,6){\vector(-4,1){3}}
\put(18,3){\vector(-4,-1){3}}
\put(18,8.5){\parbox{11\unitlength}{\small
Inhomogeneous periodic cross wire structure to create
  inhomogeneous periodic spatially dispersive medium in cavity.}}
\put(18.5,5.9){\small\  Protons}
\put(18.5,4.9){\small\ Drift tubes}
\put(18.5,5){\vector(-4,1){6}}
\put(18,3){\small\ Optimised electric field profile}

\end{picture}

\caption{Drift tube accelerator with spatially dispersive media. The fields
  experienced by the protons are optimised.}
\label{fig_drift_tube}
\end{figure}
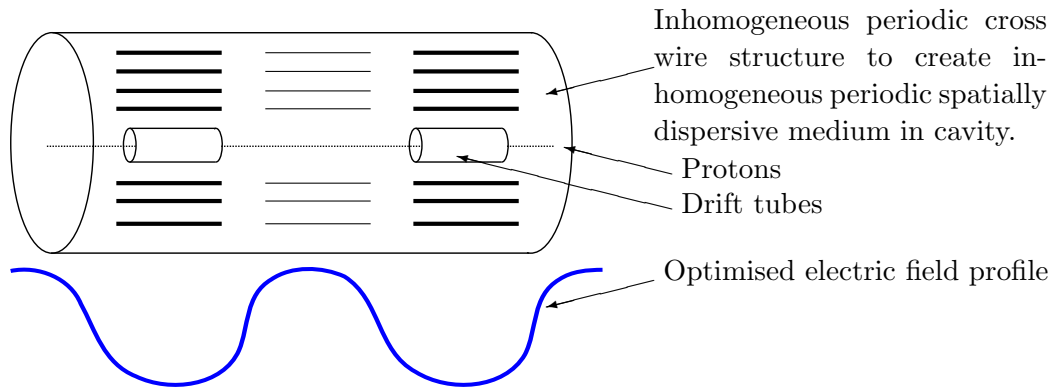

As stated in the introduction, one of the key advantages for using
inhomogeneous spatially dispersive periodic structures, is that one
can choose the shape to the propagating wave.  For example, one may
consider adding a spatially dispersive media to a drift tube in order
to optimise the fields the particles experience.  See figure
\ref{fig_drift_tube}.
Another example, in data transmission, it to shape the wave to have a
higher peak value. For example for longitudinal modes if
$\FtE(x)=-\FtP(x)
=3^{-1/2}\big(\cos(2\pi x/a) + \cos(6\pi x/a) + \cos(10\pi x/a)\big)$, then
the peak field increases by $\sqrt{3}$ but the root mean square per
length of $\FtE$ is unchanged when compared to the sinusoidal case.

\ack

The authors are grateful for the support provided by STFC (the
Cockcroft Institute ST/G008248/1), EPSRC (the Alpha-X project
EP/J018171/1) and the Lancaster University Faculty of Science and
Technology studentship program. In addition the authors would like to
thank Dr. David Burton (Physics Department, Lancaster University) and
Dr. Adam Noble (Physics Department, Strathclyde University) for help in
preparing this article. Thanks also to Prof. Alan Cairns (Physics
Department, St Andrews) for useful discussion on this subject.

\vspace{1em}

\bibliographystyle{unsrt}
\bibliography{Periodic}

\begin{thebibliography}{10}

\bibitem{costa2012macroscopic}
Jo{\~a}o~T Costa and M{\'a}rio~G Silveirinha.
\newblock Macroscopic electromagnetic response of arbitrarily shaped spatially
  dispersive bodies formed by metallic wires.
\newblock {\em Physical Review B}, 86(7):075129, 2012.

\bibitem{gratus2010covariant}
J.~Gratus and R.W. Tucker.
\newblock Covariant constitutive relations, landau damping and non-stationary
  inhomogeneous plasmas.
\newblock {\em Prog. Elec. Research M}, 13:145--156, 2010.

\bibitem{gratus2011covariant}
Jonathan Gratus and Robin~W Tucker.
\newblock Covariant constitutive relations and relativistic inhomogeneous
  plasmas.
\newblock {\em J. Math. Phys.}, 52:042901, 2011.

\bibitem{pekar1958theory}
SI~Pekar.
\newblock The theory of electromagnetic waves in a crystal in which excitons
  are produced.
\newblock {\em Soviet Journal of Experimental and Theoretical Physics}, 6:785,
  1958.

\bibitem{henneberger1998additional}
K~Henneberger.
\newblock Additional boundary conditions: An historical mistake.
\newblock {\em Physical review letters}, 80(13):2889--2892, 1998.

\bibitem{maslovski2010generalized}
Stanislav~I Maslovski, Tiago~A Morgado, M{\'a}rio~G Silveirinha, Chandra~SR
  Kaipa, and Alexander~B Yakovlev.
\newblock Generalized additional boundary conditions for wire media.
\newblock {\em New Journal of Physics}, 12(11):113047, 2010.

\bibitem{silveirinha2006additional}
M{\'a}rio~G Silveirinha.
\newblock Additional boundary condition for the wire medium.
\newblock {\em Antennas and Propagation, IEEE Transactions on},
  54(6):1766--1780, 2006.

\bibitem{silveirinha2008additional}
M{\'a}rio~G Silveirinha, Carlos~A Fernandes, and Jorge~R Costa.
\newblock Additional boundary condition for a wire medium connected to a
  metallic surface.
\newblock {\em New Journal of Physics}, 10(5):053011, 2008.

\bibitem{silveirinha2009additional}
M{\'a}rio~G Silveirinha.
\newblock Additional boundary conditions for nonconnected wire media.
\newblock {\em New Journal of Physics}, 11(11):113016, 2009.

\bibitem{agranovich1967spatial}
V~M Agranovich and V~L Ginzburg.
\newblock {\em Spatial Dispersion in Crystal Optics and the Theory of
  Excitons}.
\newblock Wiley, 1967.

\bibitem{agarwal1974electromagnetic}
GS~Agarwal, DN~Pattanayak, and E~Wolf.
\newblock Electromagnetic fields in spatially dispersive media.
\newblock {\em Physical Review B}, 10(4):1447, 1974.

\bibitem{zeyher1972spatial}
Roland Zeyher, Joseph~L Birman, and Wilhelm Brenig.
\newblock Spatial dispersion effects in resonant polariton scattering. i.
  additional boundary conditions for polarization fields.
\newblock {\em Physical Review B}, 6:4613--4616, 1972.

\bibitem{lorentz1916theory}
Hendrik~Antoon Lorentz.
\newblock {\em The Theory of Electrons: And Its Applications to the Phenomena
  of Light and Radiant Heat}, volume~29.
\newblock BG Teubner, 1916.

\bibitem{belov2003strong}
PA~Belov, R~Marques, SI~Maslovski, IS~Nefedov, M~Silveirinha, CR~Simovski, and
  SA~Tretyakov.
\newblock Strong spatial dispersion in wire media in the very large wavelength
  limit.
\newblock {\em Physical Review B}, 67(11):113103, 2003.

\bibitem{song2013accurate}
Wei Song, Zhun Yang, Xin-Qing Sheng, and Yang Hao.
\newblock Accurate modeling of high order spatial dispersion of wire medium.
\newblock {\em Optics express}, 21(24):29836--29846, 2013.

\bibitem{gratus2013Inhomogeneous}
Jonathan Gratus and Matthew McCormack.
\newblock Inhomogeneous spatially dispersive electromagnetic media.
\newblock In {\em PIERS Proceedings, Stockholm, August 12-15 2013}, pages 1007
  -- 1012. The Electromagnetics Academy, 2013.

\end{thebibliography}

\begin{appendix}
\section{}

\begin{proof}[Proof that the coefficients $P_q$ satisfy the difference equation
\eqref{Per_diff_eqn}:]
The Fourier transform of (\ref{Per_P_anzats})
\begin{equation}
\tP(k) = \sum_{q=-\infty}^\infty  P_{q} \, \delta\left(k - \left(q + \frac{\kappa}{a}\right)\right)
\label{Per_Pm_FT}
\end{equation}
Substituting the Fourier transform of (\ref{Per_L}) and
(\ref{Per_Pm_FT}) into the left hand side of
(\ref{Per_Convolution_eq}) we have the convolution
\begin{eqnarray*}
\fl&
(\tL * \tP)(k) = \int_{-\infty}^{\infty} \tL(k - k')\tP(k') dk'
\\
\fl&\quad
= \int_{-\infty}^{\infty} \!dk'\left(L_0 \delta(k-k') + \Lambda
  \delta\left(k-k' + \frac{1}{a}\right) + \Lambda \delta\left(k-k' -
  \frac{1}{a}\right)\right) \sum_{q=-\infty}^{\infty} P_{q} \delta\Big(k' -
\frac{q+\kappa}{a}\Big) 
\\
\fl&\quad
= \sum_{q=-\infty}^{\infty} P_q \left(L_0 \delta\Big(k - \frac{q +
    \kappa}{a}\Big) + \Lambda \delta\Big(k - \frac{q + \kappa + 1}{a}\Big)
  + \Lambda \delta\Big(k - \frac{q + \kappa - 1}{a}\Big)\right)
\end{eqnarray*}
by setting $q' = q-1$ and $q'' = q+1$
\begin{eqnarray*}
(\tL * \tP)(k) &= 
\sum_{q=-\infty}^{\infty} P_q L_0 \delta\Big(k - \frac{q + \kappa}{a}\Big) + 
\sum_{{q''}=-\infty}^{\infty} P_{q'' - 1} \Lambda \delta\Big(k - \frac{q''
  + \kappa}{a}\Big)  
\\&\qquad+
\sum_{{q'}=-\infty}^{\infty} P_{q' +1} \Lambda \delta\Big(k - \frac{q' + \kappa}{a}\Big)
\end{eqnarray*}
relabelling $q'$ and $q''$ as $q$
\begin{equation*}
(\tL * \tP)(k) = \sum_{q=-\infty}^{\infty} \delta\Big(k - \frac{q + \kappa}{a}\Big) (L_0 P_{q}  + \Lambda P_{q-1}  + \Lambda P_{q+1})
\end{equation*}
Equating to the right hand side of (\ref{Per_Convolution_eq})
\begin{eqnarray*}
\sum_{q=-\infty}^{\infty} 
\delta\Big(k - &\frac{q + \kappa}{a}\Big) 
(L_0 P_{q}  + \Lambda P_{q-1}  + \Lambda P_{q+1} ) 
\\&= 
\sum_{q=-\infty}^{\infty} \left(\beta^2 k^2 - 
\frac{\omega^2}{\omega^2 - k^2 c^2} \right) 
P_{q} \delta\Big(k-\frac{q + \kappa}{a}\Big)
\end{eqnarray*}
hence
\begin{eqnarray*}
\left(L_0 - \beta^2 k^2 +  \frac{\omega^2}{\omega^2 - k^2 c^2}\right)P_{q}  + \Lambda P_{q-1}  + \Lambda P_{q+1} = 0
\end{eqnarray*}
where $k = ({q + \kappa})/a$.
\end{proof}

\begin{proof}[Proof of \eqref{thmg1} and \eqref{thmg2}]
Expanding $f_n(\omega_n)$ and using (\ref{gsol1}) gives
\begin{eqnarray*}
f_{n}(\omega_n) &= 
f_n(\Omega_n)
+
(\omega_n-\Omega_{n})\FOmega'_n
+
\frac{1}{2}(\omega_n-\Omega_{n})^2
\FOmega''_n
+ \ldots
\\
&=
\Lambda^2 
\Big(\frac{1}{\FOmega_{n-1}} + \frac{1}{\FOmega_{n+1}}\Big)
+
O(\Lambda^{4})
\end{eqnarray*}
Using this and (\ref{gsolendP}) the left hand side of (\ref{thmg2})
becomes
\begin{eqnarray*}
\lefteqn{\Lambda \pn_{n-1} + f_{n}(\omega_{n})\pn_{n} + \Lambda
  \pn_{n+1}}
\qquad&
\\
&= \Lambda\left( \frac{-\Lambda}{\FOmega_{n-1}} + O(\Lambda^3)
\right) + \Lambda^2
\Big(\frac{1}{\FOmega_{n-1}} + \frac{1}{\FOmega_{n+1}}\Big)
+ O(\Lambda^4) 
\\&\qquad
+  \Lambda \left( \frac{-\Lambda}{\FOmega_{n+1}} +
  O(\Lambda^3) \right)
\\
  &= -\left( \frac{\Lambda^2}{\FOmega_{n-1}} +
    \frac{\Lambda^2}{\FOmega_{n+1}} \right)
+
\left( \frac{\Lambda^2 }{\FOmega_{n-1}} + \frac{\Lambda^2 }{\FOmega_{n+1}}
\right)  + O(\Lambda^4)
\\\fl
  &= O(\Lambda^4)
\end{eqnarray*}
hence \eqref{thmg2}. To prove (\ref{thmg1}) consider first the case when
$q > n$, then using $f_q(\omega_n) = \FOmega_q+ O(\Lambda^2)$ we get
\begin{eqnarray*}
\fl
\lefteqn{\Lambda \pn_{q-1} + f_{q}(\omega_n) \pn_{q} + \Lambda \pn_{q +1}
= \Lambda \left(\frac{(-\Lambda)^{q-1-n}}{\prod_{k=1}^{q-1-n}
    \FOmega_{n+k}} + O(\Lambda^{q-n + 1}) \right)
 }&
\\\fl
&\ \ \ + f_{q}(\omega_n)
\left(\frac{(-\Lambda)^{q-n}}{\prod_{k=1}^{q-n} \FOmega_{n+k}} +
  O(\Lambda^{q-n + 2}) \right)
+ \Lambda \left(\frac{(-\Lambda)^{q+1-n}}{\prod_{k=1}^{q+1-n}
    \FOmega_{n+k}} + O(\Lambda^{q-n + 3}) \right)
\\\fl&= 
\frac{(-\Lambda)^{q-1-n}}{\prod_{k=1}^{q-1-n}\FOmega_{n+k}}\left(\Lambda - f_{q}(\omega_n)\frac{\Lambda}{\FOmega_{q}} \right) + O(\Lambda^{q-n+2})
\\\fl&= 
\frac{(-\Lambda)^{q-1-n}}{\prod_{k=1}^{q-1-n}\FOmega_{n+k}}
\left(\Lambda - \left(f_q(\Omega_n) + O(\Lambda^2)\right)
\frac{\Lambda}{\FOmega_{q}} \right) + O(\Lambda^{q-n+2})
\\\fl&= 
\frac{(-\Lambda)^{q-1-n}}{\prod_{k=1}^{q-1-n}\FOmega_{n+k}}\left(\Lambda - \Lambda \right) + O(\Lambda^{q-n+2})\\ &= O(\Lambda^{q-n+2})
\end{eqnarray*}
The case for $q < n$ follows likewise.
\end{proof}

\end{appendix}

\end{document}